\documentclass[showpacs,twocolumn,pra,secnumeric,amsmath,amssymb,letterpaper]{revtex4}

\usepackage{graphicx}% Include figure files
\usepackage{dcolumn}% Align table columns on decimal point
\usepackage{bm}% bold math

\begin{document}

\title{Intensity correlations and entanglement by frequency doubling\\in a dual ported resonator}

\author{O.-K. Lim}
\email{oklim@wisc.edu}
\author{M. Saffman}
\affiliation{
Department of Physics,
University of Wisconsin,
1150 University Avenue, 
 Madison, WI 53706.
}
\date{\today}

\begin{abstract}
We show that non-classical intensity correlations and quadrature entanglement can be generated by frequency doubling in a resonator with two output ports. We predict  twin-beam intensity correlations  6 dB below the coherent state limit, and that the product of the inference variances of the quadrature fluctuations gives an Einstein-Podolsky-Rosen (EPR) correlation coefficient of $V_{\rm EPR}=0.6<1$. 
Comparison with an entanglement source based on combining two frequency doublers with a beam splitter shows that the dual ported resonator provides stronger entanglement at lower levels of individual beam squeezing. Calculations are performed using a self-consistent propagation method that does not invoke a mean field approximation.
Results are given for physically realistic parameters that account for 
the Gaussian shape of the intracavity beams, as well as intracavity losses.  
\end{abstract}

\maketitle

\section{\label{sec:level1}Introduction}
The process of parametric down conversion has been used widely to generate non-classical optical fields  at the level of single photons, as well as many photon fields described by continuous variables\cite{Drummondbook}. At the microscopic level non-classical correlations and entanglement arise due to the possibility of converting a single high frequency photon into a pair of correlated lower frequency photons. 
The down conversion process can be used to generate so called twin beams that have intensity correlations that are stronger than would be obtained with individual coherent states\cite{Heidmann87, Tapster88, Aytur90, Peng98, Gao98, Zhang00, Hayasaka04}. Twin beams have been successfully applied to sub-shot-noise spectroscopy~\cite{Ribeiro97} and quantum nondemolition measurements~\cite{Schiller96}.  
The generation of entangled beams in parametric down conversion was predicted by Reid and Drummond in the late 
80's\cite{Reid88,Reid89,Drummond90}, and is crucial for studies of  quantum teleportation and networking\cite{Furusawa98,Schori02,Bowen03,Jia05,Koike06}.

The process of second harmonic generation (SHG) where a fundamental field at  $\omega_1$ is frequency doubled to create a harmonic at frequency $\omega_2=2\omega_1$ is complementary to parametric down conversion, and can also be used for creating nonclassical light. The possibility of using frequency doubling may be convenient for reaching spectral regions that are not readily accessible by down conversion.  It is well known that quadrature squeezing of both the fundamental and harmonic fields occurs in second harmonic generation\cite{Pereira88,Sizmann90,Paschotta94}.
The generation of multibeam correlations in second harmonic generation is less well studied than in the case of parametric down conversion.
Calculations have demonstrated the existence of correlations between the fundamental and harmonic fields 
 \cite{Horowicz89,Dance93,Wiseman95,Olsen99}  including  entanglement between the fundamental and harmonic fields\cite{Olsen04,Grosse06} and entanglement in type II SHG in the fundamental fields alone\cite{Andersen03a,Andersen03b}. The possibility of nonclassical spatial correlations in either the fundamental or harmonic fields alone\cite{Lodahl02,Lodahl02e,Bache02} and of entanglement   in the fundamental field\cite{Lodahl03} has also been shown in models that include diffraction.

\begin{figure}
\centering
\includegraphics[scale=.4]{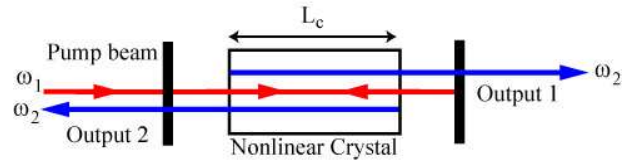}
\caption{\label{fig:linear}(color online) A dual ported singly resonant cavity which provides two harmonic outputs.}
\end{figure}

In this work we investigate the production of two beams at the harmonic frequency that exhibit nonclassical intensity correlations and quadrature entanglement. The device we analyze is the dual ported resonator shown in Fig. \ref{fig:linear}. The pump beam at $\omega_1$ is resonant in the cavity, while the generated harmonic exits at both end mirrors. As we show below the harmonic outputs exhibit strong quantum correlations. This can be understood naively as follows. While traversing the crystal to the right squeezing is generated in the fundamental and harmonic fields, as well as correlations between the fundamental and harmonic. Even though all of the harmonic leaves the cavity at the right hand mirror, the intracavity fundamental that generates a harmonic field on the backwards pass through the crystal is correlated with harmonic output 1.  The intracavity field transfers the correlations to output 2, leading to a nonzero correlation between outputs 1 and 2. 

In order to describe this process theoretically we need to go beyond the usual mean field description of SHG squeezing\cite{Paschotta94}, and account for variations in the fields at different locations inside the cavity. To do so we combine the linearized solutions for propagation of quantum fluctuations in traveling wave SHG\cite{Ou94,Li94}
with a self-consistent application of the cavity boundary conditions as was first done by Maeda and Kikuchi\cite{Maeda97}. It was shown in Ref. \onlinecite{Maeda97} that the propagation model reproduces the mean field results at low coupling strength, but predicts larger quantum noise reduction at high coupling strength. Since the propagation model relies on a linearized description of the quantum fluctuations its validity may break down in the limit of very large squeezing, where the fluctuations in the unsqueezed quadrature are no longer small. The accuracy of the linearized model was studied by comparison with numerical solutions of a quantum phase space model\cite{Olsen99}. It was found that the linearized model provides accurate predictions provided the normalized interaction length $\zeta$\cite{Ou94,Li94} does not exceed 2-3. In the results presented here, using realistic physical parameters and Gaussian beams, we have $\zeta<1$. We are therefore confident that the linearized theory provides an accurate description of the nonlinear resonator for the range of parameters considered here.  

The structure of the paper is as follows. In Sec.  \ref{sec:SHG} we define notation and present the solutions for the quadrature fluctuations and squeezing spectra of the harmonic outputs.  These results are used to calculate the intensity correlations in 
Sec. \ref{sec:Twin}. EPR correlations and inseparability\cite{Duan00} of the outputs are demonstrated in Sec. \ref{sec:EPR}. We compare the EPR correlations created in the dual ported resonator with the alternative approach of combining two separate SHG resonators at a beamsplitter in Sec. \ref{sec:compare}, and give a discussion of the results obtained in Sec. \ref{sec:discussion}.

\section{ Propagation model of intracavity SHG}
\label{sec:SHG}

\begin{figure}[!t]
\begin{center}
\includegraphics[width=8.cm]{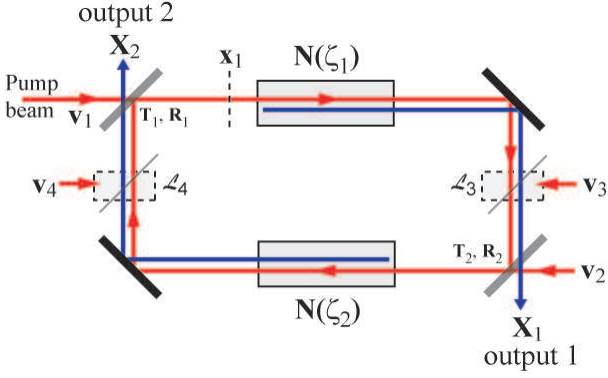}
\caption{(color online) Propagation model of singly resonant SHG cavity with dual output ports.}
\label{fig:unfolded}
\end{center}
\end{figure}

The dual ported resonator of Fig. \ref{fig:linear} is shown in more detail in Fig. \ref{fig:unfolded} where we have separated the forward and backward passes to create an equivalent ring resonator model. Since the  counterpropagating beams are strongly phase mismatched in the crystal we can neglect any direct interaction between them. 
The pump beam at frequency $\omega_1$ enters through an input coupler described by transmission and reflection matrices ${\bf T}_1, {\bf R}_1$ (defined below).  The second harmonic outputs are allowed to escape through input and output mirrors. The cavity is assumed to be resonant at the frequency $\omega_{1}$ of the external pump beam.

%\subsection{\label{sec:mean}Mean field solutions for quantum noise fluctuations in single pass SHG}

The coupled propagation of the slowly varying wave envelopes of the fundamental (frequency $\omega_1$) and second harmonic (frequency $\omega_2$)  in the crystal is described by the classical equations
\begin{subequations}
\begin{align}
\frac{\partial {\mathcal E}_1}{\partial z}&=i\kappa_1 {\mathcal E}_1^*{\mathcal E}_2 e^{-i\Delta kz},\\
\frac{\partial {\mathcal E}_2}{\partial z}&=i\kappa_2 {\mathcal E}_1^2 e^{i\Delta kz}.
\end{align}
\end{subequations}
The intensities of each field are given by $I_{i}$=$\frac{\epsilon_{0}}{2}n_{i}c|{\mathcal E}_{i}|^2$, $n_i$ is the refractive index at frequency $\omega_i$, $c$ is the speed of light in vacuum, and the phase mismatch is $\Delta k=2k_1- k_2$ where $k_i=\omega_i n_i/c.$
The coupling constants are given by $\kappa_i=\omega_1 d/(c n_i),$
where $d$ is the effective second-order susceptibility of the crystal. 

These equations are valid for plane waves whereas real experiments are typically performed with Gaussian beams. In order to obtain the correct value for the coupling constant when using Gaussian beams we avail ourselves of the results of the Boyd-Kleinman theory\cite{Boyd68}. We assume that the fundamental field is a lowest order Gaussian beam with  waist $w$ ($1/e^2$ intensity radius) that is symmetrically located in the center of the crystal and  
 introduce scaled field amplitudes $A_i=i \sqrt{\epsilon_o c n_i/(2\hbar \omega_i)} {\mathcal E}_i$. The characteristic field strength ${\mathcal E}_i$ is chosen such that $|A_i|^2$ gives the number of photons per second at frequency $\omega_i$ carried by the Gaussian beam.
The coupled equations for the scaled amplitudes then take the form
\begin{subequations}
\begin{align}
\frac{\partial A_1}{\partial z}&=-\kappa^* A_1^*A_2,\\
\frac{\partial A_2}{\partial z}&=\frac{\kappa}{2} A_1^2,
\end{align}
\label{eq.dAdz}
\end{subequations}
where the complex coupling coefficient is 
\begin{subequations}
\begin{align}
\kappa &=\left(\frac{2 n_1 \hbar \omega_1 E_{NL}}{ n_2 L_c^2}\right)^{1/2}e^{\imath\phi_h}\\
E_{NL}&=\frac{2\omega_1^2 d^2}{\epsilon_0 c^3 n_1^2 n_2 }\frac{L_c^2}{\pi w^2}| h|^2~~~\\
h &= \int_{-1/2}^{1/2} d\xi\, \frac{e^{\imath \Delta k L_c \xi}}{1+i\frac{L_c }{z_{R1}}\xi},
\end{align}
\label{eq.kappa}
\end{subequations}
where  $\phi_h=\arg(h),$ $L_c$ is the length of the crystal, and 
$z_{R1}= \pi n_1 \omega_1 w^2/(2\pi c)$ is the Rayleigh length of the fundamental beam.  The single pass conversion efficiency is determined by $E_{NL}=\frac{P_2(z=L_c)}{P_1^2(z=0)}$, where 
$P_1, P_2$ are the powers of the two fields. The Boyd-Kleinman solution assumes that the harmonic is generated with a waist that gives equal Rayleigh lengths for both frequencies, and that the attenuation of the fundamental field due to upconversion is small. As shown in Appendix \ref{sec:appA} the fractional depletion of the intracavity field due to a single pass through the crystal is always small for the parameters considered here.  For parameters where this is  not the case it would be necessary to use a more cumbersome multimode theory that accounts for the coupling of different spatial modes\cite{Schwob98} which is outside  the scope of the present work.

To solve for the transformation of quantum fluctuations we replace the classical fields in Eqs. (\ref{eq.dAdz}) by annihilation operators $\hat A_j.$ The propagation equations for $\hat A_j$ are the same as the classical Eqs. (\ref{eq.dAdz}). These nonlinear operator equations can be solved perturbatively by putting $\hat A_j(z,t)= A_j(z) +\hat a_j(z,t)$
where $A_j(z)$ are the (classical) mean fields and $\hat a_j(z,t)$ time and space dependent  fluctuation operators which satisfy the commutation relations 
$[\hat a_i(z,t),\hat a_j^\dag(z',t')]=\delta_{ij}\delta(z-z')\delta(t-t'),$ $[\hat a_i(z,t),\hat a_j(z',t')]=0.$ We then linearize in the fluctuation operators, and write the resulting operator equations as equations for classical fluctuations with the replacements $\hat a_j(z,t)\rightarrow a_j(z,t)$, $\hat a_j^\dag(z,t)\rightarrow a_j^*(z,t)$.  The semiclassical theory\cite{Reynaud89} expresses the expectation values of symmetrically ordered quantum operators in terms of the classical   
c-numbers $a_j, a_j^*$.

The solutions to the linearized propagation equations are known\cite{Ou94,Li94}. We can write the solutions in the form ${\bf x}(\zeta,t)={\bf N}(\zeta) {\bf x}(0,t)$ where ${\bf x}=(x_1,x_2,y_1,y_2)^T$ and
\begin{eqnarray}
 x_j(z,t)&=& a_j(z,t) + a_j^*(z,t),\nonumber\\
 y_j(z,t)&=&-i[ a_j(z,t)- a_j^*(z,t)],\nonumber
\end{eqnarray}
are the amplitude and phase quadrature fluctuations. When $A_2(0)=0$ the transformation matrix is given by 
\begin{equation}
{\bf N}=\left(\begin{matrix}
N_{11}&N_{12}&0&0\\
N_{21}&N_{22}&0&0\\
0&0&N_{33}&N_{34}\\
0&0&N_{43}&N_{44}
\end{matrix}
\right).
\label{eq.N}
\end{equation}
Expressions for the matrix elements $N_{ij}(\zeta)$ are given in Appendix \ref{sec:appB}.
The normalized propagation length is given by 
\begin{equation}
\zeta=\frac{1}{\sqrt2}|A_1(0)||\kappa| L_c.
\end{equation}
Using Eqs. (\ref{eq.kappa}) we can express $\zeta$ in terms of experimentally accessible parameters as 
\begin{equation}
\zeta=\left(\frac{n_1 P_1(0)E_{NL}}{n_2} \right)^{1/2}.
\end{equation}

At this point we note that the transformation matrix (\ref{eq.N}) is only valid when $\kappa$ is real. In the more general case of complex $\kappa$ propagation mixes the $x$ and $y$ quadratures. We will limit our study to the case of real $\kappa$ for which the analytical solutions can be expressed in terms of simple hyperbolic functions (the more general case involves elliptic functions). We therefore wish to have $\kappa$ real which is the case when $\Delta k=0.$ Unfortunately the maximum value of $E_{NL}$ and hence $\zeta$ is obtained for\cite{Boyd68}  $z_{R1}=0.176 L_c$ and 
$\Delta k L_c=3.26$ which implies a complex value of $\kappa.$ For the analysis presented below we choose $\Delta k = 0$ in order to make $\kappa$ real. For zero phase mismatch the optimum value of the beam focusing corresponds to $z_{R1}=0.325 L_c$, which gives a value of $E_{NL}$ that is about 40\% smaller than could be obtained with nonzero phase mismatch. We will assume these focusing conditions in all the subsequent analysis. Numerical results will be given for a KNbO$_3$ crystal, fundamental wavelength of $\lambda_1=860~\rm nm$, $L_c=1~\rm cm,$ $d=11~\rm pm/V,$ $n_1=n_2=2.2,$  and $\Delta k=0$ for which the optimum focusing is $w=21.1~\mu\rm m$ which gives $E_{NL}=0.015~\rm W^{-1}.$ 

A self-consistent solution for the fluctuations in the cavity of Fig. \ref{fig:unfolded} is found by combining the transfer matrix for the crystal propagation with the effects of mirror reflections and transmissions as well as intracavity losses. We first transform to frequency domain variables defined by 
\begin{subequations}
\begin{eqnarray}
&&a_j(z,\Omega)=\int dt\, a_j(z,t)e^{\imath\Omega t}\\
&&a_j^*(z,-\Omega)\equiv [a_j(z,-\Omega)]^*=\int dt\, a_j^*(z,t)e^{\imath\Omega t}.
\end{eqnarray}
\end{subequations}
The corresponding frequency domain quadrature fluctuations are 
\begin{subequations}
\begin{eqnarray}
x_j(z,\Omega)&=& a_j(z,\Omega) + a_j^*(z,-\Omega ),\\
 y_j(z,\Omega)&=&-i[ a_j(z,\Omega)- a_j^*(z,-\Omega)].
\end{eqnarray}
\label{eq.xyomega}
\end{subequations}
Note that in the frequency domain the quadrature amplitudes $x(z,\Omega),y(z,\Omega)$ are complex variables. 

We introduce the  $4\times 4$  matrices for transmission and reflection 
with nonzero diagonal elements ${\mathbf T}_j={\rm diag}(\sqrt{T_{1j}},\sqrt{T_{2j}},\sqrt{T_{1j}},\sqrt{T_{2j}})$,
  ${\mathbf R}_j={\rm diag}(\sqrt{1-T_{1j}},\sqrt{1-T_{2j}},\sqrt{1-T_{1j}},\sqrt{1-T_{2j}})$, 
where $T_{ij}$ denotes the power transmittance for frequency $i$ at mirror $j$. 
Residual intracavity losses due to crystal absorption, reflections at crystal surfaces, and mirror losses are lumped into effective loss beamsplitters indicated by ${\mathcal L}_3,{\mathcal L}_4$ in Fig. \ref{fig:unfolded}. These losses are described by coefficients $L_{ij}$ for frequency $\omega_i$ at position $j$  and corresponding reflection and transmission matrices
${\mathbf T}_{{\mathcal L}_j}={\rm diag}(\sqrt{1-L_{1j}},\sqrt{1-L_{2j}},\sqrt{1-L_{1j}},\sqrt{1-L_{2j}})$,
${\mathbf R}_{{\mathcal L}_j}={\rm diag}(\sqrt{L_{1j}},\sqrt{L_{2j}},\sqrt{L_{1j}},\sqrt{L_{2j}})$.

The phase shift acquired in one cavity round trip is represented by 
the matrix ${\mathbf D}={\rm diag}(e^{\imath \Omega/\nu_{c1}},e^{\imath  \Omega/\nu_{c2}} ,e^{\imath   \Omega/\nu_{c1}},e^{\imath   \Omega/\nu_{c2}})$,
where the cavity free spectral range is $\nu_{ci}=\ c/(2n_i L_c + 2L_a)$, and $2L_a$ is the round trip length of air in the cavity.
 We have assumed the cavity is on resonance, so that the phase shift is an odd function of $\Omega.$ This ensures that $\bf D$  can be used with the quadrature fluctuation vector ${\bf x}(z,\Omega)$ which contains components at $\pm\Omega.$ If the cavity were detuned it would mix the $x, y$ quadratures and the round trip phase  would have to be calculated separately for $a(z,\Omega)$ and $a^*(z,-\Omega).$

Vacuum noise sources enter the cavity at mirrors $1,2$ and through the loss ports. We describe these by quadrature noise vectors
${\mathbf v}_{j}=(u_{1j}(\Omega),u_{2j}(\Omega),v_{1j}(\Omega),v_{2j}(\Omega))^T,$ where $u_{ij}(\Omega), v_{ij}(\Omega)$ are frequency domain amplitude and phase quadrature fluctuations of frequency $\omega_i$ at position $j.$  

Using Eqs. (\ref{eq.N}-\ref{eq.xyomega}) and the above definitions the self-consistent solution for the intracavity fluctuations to the right of beamsplitter $1$ is 
 \begin{widetext}
\begin{eqnarray}
{\bf x}_1 &=&[{\bf I}-{\bf D}{\bf R}_1{\bf T}_{{\mathcal L}_4} {\bf N}(\zeta_2){\bf R}_2{\bf T}_{{\mathcal L}_3}{\bf N}(\zeta_1)]^{-1}{\bf D}[{\bf T}_1{\bf v}_1+{\bf R}_1{\bf T}_{{\mathcal L}_4} {\bf N}(\zeta_2){\bf T}_2{\bf v}_2-{\bf R}_1 {\bf T}_{{\mathcal L}_4} {\bf N}(\zeta_2){\bf R}_2 {\bf R}_{{\mathcal L}_3}{\bf v}_3 -  {\bf R}_1 {\bf R}_{{\mathcal L}_4}{\bf v}_4]\nonumber\\
\end{eqnarray}
where $\bf I$ is the identity matrix. The result depends on the propagation lengths $\zeta_1$ and $\zeta_2$ which in turn are functions of the pump beam power and resonator parameters.  Expressions for the propagation lengths in terms of experimentally accessible parameters are given in Appendix \ref{sec:appA}.
The vectors of 
output quadrature fluctuations 
are defined as ${\bf X}_j=(X_{1j},X_{2j},Y_{1j},Y_{2j})^T$, where 
$j$ labels the spatial position.
The outputs can be written in terms of the self-consistent intracavity field as 
\begin{subequations}
\begin{align}
{\bf X}_1&={\bf T}_2 {\bf T}_{{\mathcal L}_3}  {\bf N}(\zeta_1){\bf x}_1-{\bf R}_2{\bf v}_2-{\bf T}_2 {\bf R}_{{\mathcal L}_3}
{\bf v}_3 \\
{\bf X}_2&={\bf T}_1{\bf T}_{{\mathcal L}_4}  {\bf N}(\zeta_2){\bf R}_2{\bf T}_{{\mathcal L}_3}  {\bf N}(\zeta_1){\bf x}_1-{\bf R}_1{\bf v}_1+{\bf T}_1{\bf T}_{{\mathcal L}_4} {\bf N}(\zeta_2){\bf T}_2{\bf v}_2-{\bf T}_1{\bf T}_{{\mathcal L}_4} {\bf N}(\zeta_2){\bf R}_2{\bf R}_{{\mathcal L}_3}{\bf v}_3-{\bf T}_1{\bf R}_{{\mathcal L}_4}{\bf v}_4.
\end{align}
\label{eq.outputs}
\end{subequations}
\end{widetext}

Equations (\ref{eq.outputs}) are quite general and can be used to describe singly or doubly resonant cavities provided  the nonlinear propagation is phase matched and the cavity is on resonance at both frequencies. The general expressions for the output quadratures that result from evaluation of these equations are very cumbersome. We will restrict ourselves to the case of a resonant fundamental, and complete transmission of the harmonic at mirrors $1, 2$, i.e. $T_{21}=T_{22}=1.$ We will assume that the intracavity losses ${\mathcal L}_3, {\mathcal L}_4$ only affect the fundamental fields so $L_{23}=L_{24}=0.$ This last assumption is not a loss of generality since harmonic losses can be accounted for  at the detectors external to the cavity. Finally, since we will take $L_{13}\ne 0$ which accounts for fundamental loss between the two passes through the crystal, we can put $T_{12}=0$ without loss of generality. With these assumptions the results for the harmonic output quadrature fluctuations can be written as 
\begin{subequations}
\begin{eqnarray}
X_{21}&=& f_{11}u_{11}+f_{13}u_{13}+f_{14}u_{14}+f_{21}u_{21}+f_{22}u_{22}\nonumber\\
&&\\
Y_{21}&=&g_{11}v_{11}+g_{13}v_{13}+g_{14}v_{14}+g_{21}v_{21}+g_{22}v_{22}
\nonumber\\
&&\\
X_{22}&=& h_{11}u_{11}+h_{13}u_{13}+h_{14}u_{14}+h_{21}u_{21}+h_{22}u_{22}\nonumber\\
&&\\
Y_{22}&=&j_{11}v_{11}+j_{13}v_{13}+j_{14}v_{14}+j_{21}v_{21}+j_{22}v_{22}.\nonumber\\
\end{eqnarray}
\label{eq.outputs2}
\end{subequations}
Explicit expressions for the coefficients $f,g,h,j$ are given in Appendix \ref{sec:appB}. 
Except when needed for clarity we will in what follows suppress the  dependence on $z$ and $\Omega$ for brevity.

\begin{figure*}[!t]
\begin{center}
\includegraphics[width=.9\textwidth]{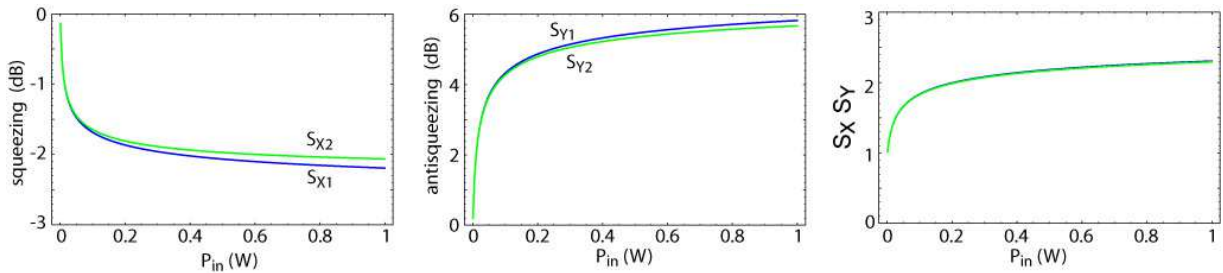}
\caption{(color online) Amplitude squeezing (left) and phase antisqueezing (center) at $\Omega=0$ for the parameters given in the text. The right hand plot shows the uncertainty product which is almost identical for the two output ports.}
\label{fig:squeeze}
\end{center}
\end{figure*}

We can use the solutions (\ref{eq.outputs2}) to calculate the  normalized squeezing spectra of the harmonic fields at output port $j$ defined by
\begin{subequations}
\begin{eqnarray}
S_{Xj}&=&\frac{\langle X_{2j}(\Omega)X_{2j}^*(\Omega)\rangle}{\langle u_{21}(\Omega)u_{21}^*(\Omega)\rangle},\\
S_{Yj}&=&\frac{\langle Y_{2j}(\Omega)Y_{2j}^*(\Omega)\rangle}{\langle v_{21}(\Omega)v_{21}^*(\Omega)\rangle}.
\end{eqnarray}
\label{eq.spectra}
\end{subequations}
The spectra $S_{Xj}, S_{Yj}$ are normalized by the input noise so that
$S<1$ corresponds to a squeezed quadrature. 
To evaluate the spectra we make the usual assumption that the input noise fields at different locations and frequencies are uncorrelated so that 
\begin{eqnarray}
<u_{ij}(\Omega) u_{kl}^*(\Omega')> &=&<v_{ij}(\Omega) v_{kl}^*(\Omega')> = \delta_{ik}\delta_{jl}\delta(\Omega-\Omega').
\nonumber\\
%\\
% <u_{ij}(\Omega) u_{kl}(\Omega')> &=&<v_{ij}(\Omega) v_{kl}(\Omega')> = 0,\\
% <u_{ij}(\Omega) v_{kl}(\Omega')> &=&<u_{ij}(\Omega) v_{kl}^*(\Omega')> = 0.
\label{eq.noise}
\end{eqnarray}
Using Eq. (\ref{eq.noise}) we can write the output squeezing spectra of the harmonic fields as
\begin{subequations}
\begin{eqnarray}
S_{X1}&=& |f_{11}|^2+|f_{13}|^2+|f_{14}|^2+|f_{21}|^2+|f_{22}|^2,\nonumber\\
&&\\
S_{Y1}&=&|g_{11}|^2+|g_{13}|^2+|g_{14}|^2+|g_{21}|^2+|g_{22}|^2,
\nonumber\\
&&\\
S_{X2}&=& |h_{11}|^2+|h_{13}|^2+|h_{14}|^2+|h_{21}|^2+|h_{22}|^2,\nonumber\\ 
&&\\
S_{Y2}&=&|j_{11}|^2+|j_{13}|^2+|j_{14}|^2+|j_{21}|^2+|j_{22}|^2.\nonumber\\
\end{eqnarray}
\label{eq.spectra2}
\end{subequations}
It can be readily verified using the expressions given in Appendix \ref{sec:appB} that when $\zeta_1=\zeta_2=0$ which corresponds to a purely linear 
resonator all the squeezing spectra are identically unity.

Representative results for the squeezing spectra are given in Fig. \ref{fig:squeeze}. The results are shown as a function of the pump beam power external to the cavity for parameters $E_{NL1}=E_{NL2}=.015~\rm W^{-1},$
$T_{11}=.01,$ and $L_{13}=L_{14}=0.005.$ The dependence of $\zeta_1, \zeta_2$ on pump power is shown in Appendix \ref{sec:appA}. The smaller squeezing effect on the second output beam can be attributed to the fact that the mean field value of the intracavity power is reduced due to the SHG process after the forward pass through the crystal. Since, as is shown in Appendix \ref{sec:appA}, the normalized propagation lengths are less than  0.25 at the highest pump power used, we can expect the linearized semiclassical results to be 
accurate\cite{Olsen99}.

\section{\label{sec:Twin}Nonclassical intensity correlations }

Given the squeezing spectra we can evaluate  the quantum correlation of the intensity difference or sum of the two output beams. 
Using two detectors we measure the intensity difference of the harmonic outputs as
$$
I_-=I_1-g I_2
$$
where $g$ is an electronic  gain factor. As the intensities of the two harmonic outputs are not equal due to different propagation lengths $\zeta_1, \zeta_2$ as well as the possibility 
of unequal  detector sensitivities we introduce an electronic gain parameter $g$ that can be adjusted to minimize the noise of the intensity difference or sum. 
We can express the variance of the detected intensity difference fluctuations in terms of the squeezing spectra calculated above as follows. The fluctuations of a beam with intensity $I$
are $ \delta I(\Omega) = \sqrt{2I}X(\Omega).$  The variance of the fluctuations is given by
\begin{eqnarray}
(\Delta|\delta I|)^2&=&\langle |\delta I|^2\rangle-|\langle \delta I\rangle|^2\nonumber\\
&=& 2 I \langle |X(\Omega)|^2 \rangle.
\label{eq.diffvariance1}
\end{eqnarray}
We have normalized the fields such that for a coherent state 
$S_X^{\rm coh}(\Omega)=\langle |X^{\rm coh}(\Omega)|^2 \rangle/
\langle |u(\Omega)|^2 \rangle=1,$ thus the normalized variance of the detected signal due to a coherent state with  intensity $I$ is just $2I.$

The corresponding formula for the intensity difference fluctuations is  
$$
\delta I_- =  \sqrt{2 I_1} X_{21}- g\sqrt{2 I_2} X_{22}.
$$
The variance of the 
intensity difference fluctuations is thus
\begin{eqnarray}
(\Delta|\delta I_-|)^2&=&\langle |\delta I_-|^2\rangle-|\langle \delta I_-\rangle|^2\nonumber\\
&=& 2 I_1 \langle |X_{21}(\Omega)|^2\rangle  +g^2 2 I_2 \langle |X_{22}(\Omega)|^2\rangle \nonumber\\
&&- 2 g \sqrt{I_1 I_2 }\tilde C_X ,\nonumber
\label{eq.diffvariance}
\end{eqnarray}
where we have introduced the correlation coefficient 
$\tilde C_{X}=\langle  X_{21}(\Omega){X_{22}}^*(\Omega)+
{X_{21}}^*(\Omega)X_{22}(\Omega) \rangle.$
Normalizing by the sum of the shot noise variance for coherent state outputs with the same total intensity $ (2 I_1  + g^2 2I_2 ) \delta(0) $ we obtain 
\begin{eqnarray}
(\Delta|\delta I_-|)_{\rm norm}^2
&=& \frac{I_1 S_{X1} + g^2 I_2 S_{X2}}{I_1+ g^2 I_2}-g\frac{  \sqrt{I_1  I_2 }C_{X}}{ I_{1} + g^2  I_{2} } ,
\label{eq.diffvariance2}
\end{eqnarray}
where $C_X=\tilde C_X/\delta(0).$
When the second term is negative the variance can be less than unity which represents a nonclassical twin beam correlation.

The optimum value of $g$ which minimizes the fluctuations is 
found by putting $\partial (\Delta|\delta I_-|)_{\rm norm}^2/\partial g=0$ which gives 
$$
g_{\rm opt}=\sqrt{\frac{I_1 }{I_2  } }
\frac{ S_{X1}-S_{X2}\pm \sqrt{(S_{X1}-S_{X2})^2+C_X^2}}{C_X} .
$$
As shown in Appendix \ref{sec:appA} the intensity ratio is given by $I_1/I_2=\epsilon_1/\epsilon_2$ which can be determined from Eqs. (\ref{eq.eps1}-\ref{eq.eps2}).
For input powers up to a few Watts $I_1\simeq I_2$ and $S_{x1}\simeq S_{x2}.$ Thus the optimum g values are $g_{\rm opt}\simeq \pm 1.$

 The case of $g\simeq 1$ corresponds to an intensity difference which has noise greater than the shot noise limit, while  $g\simeq -1$ which corresponds to the sum of intensities gives a strong nonclassical correlation. There is a simple physical explanation of this effect. 
A positive amplitude fluctuation in the harmonic output at port 1 corresponds to an increased depletion of the fundamental beam. The weakened fundamental then results in a smaller amplitude output of the harmonic at the second port. Thus the correlation function $C_X$  is negative and the sum of the output intensities has a reduced expectation value. 

\begin{figure}[!t]
\begin{center}
\includegraphics[width=8.cm]{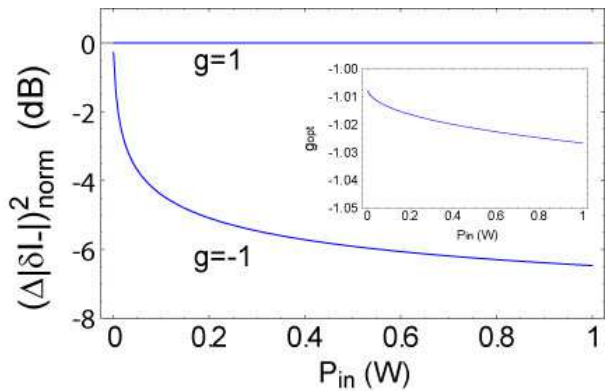}
\caption{(color online) Normalized fluctuations of the difference and sum  intensities at $\Omega=0$ using $g=\pm 1.$ The inset shows $g_{\rm opt}$ for the intensity sum. The parameters used were the same as in Fig. \ref{fig:squeeze}.}
\label{fig:Idiff}
\end{center}
\end{figure}

Using the optimum values $g_{\rm opt}$ the normalized variance given by Eq. (\ref{eq.diffvariance2}) can be written as 
\begin{eqnarray}
(\Delta|\delta I_-|)_{\rm norm}^2&=& \frac{S_{X1}+S_{X2}\pm  \sqrt{
(S_{X1}-S_{X2})^2 +C_X^2}}
{2 },\nonumber\\
\label{eq.diffvariance3}
\end{eqnarray}
where the minus sign corresponds to the case $g\simeq -1.$ The normalized intensity noise is shown in Fig. \ref{fig:Idiff} as a function of pump power. The fluctuations of the intensity sum for $g=g_{\rm opt}$ are indistinguishable from the case $g=-1$ for the range of pump powers shown.  
We see that the nonclassical intensity correlation is stronger than the 
squeezing of each output beam shown in Fig. \ref{fig:squeeze}.

\section{\label{sec:EPR}EPR Correlations and entanglement}
%\subsection{Dual ported cavity}

\begin{figure}[!t]
\begin{center}
\includegraphics[width=8.cm]{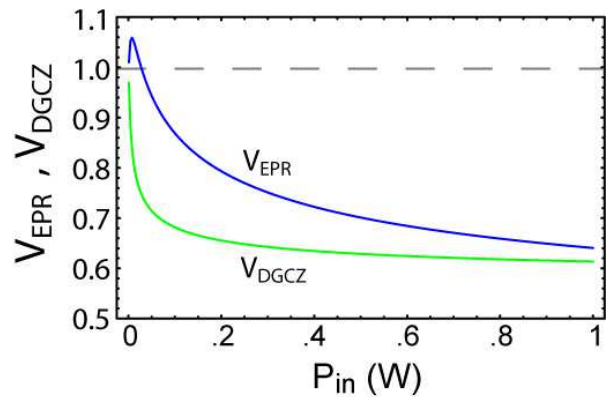}
\caption{(color online) Normalized inferred variance of the harmonic outputs at $\Omega=0.$  The parameters used were the same as in Fig. \ref{fig:squeeze}. }
\label{fig.epr1}
\end{center}
\end{figure}

The presence of nonclassical twin beam correlations motivates evaluating the presence of EPR correlations and entanglement between the two output beams. Optical beams with quadrature fluctuations that embody  EPR correlations were first demonstrated in 1992\cite{Ou92}. 
The essence of the EPR paradox is the ability to infer an observable of one system from a measurement performed on a second system spatially separated from the first.   Hence  a conditional variance is used to quantify the degree of EPR correlation. As shown by Reid\cite{Reid89} a linear estimate of the inference variance  can be used as a sufficient condition for the presence of the EPR paradox.  We  define the  normalized inference variances as 
\begin{eqnarray}
(\Delta X)_{\rm inf}^2&=& \langle |X_{21}-g_X X_{22}|^2\rangle/\delta(0)\nonumber\\
(\Delta Y)_{\rm inf}^2&=& \langle |Y_{21}-g_Y Y_{22}|^2\rangle/\delta(0).
\label{eq.epr1}
\end{eqnarray}
Here $g_X, g_Y$ are real gain parameters that are chosen to minimize the  inference variances. The condition for EPR correlations is 
$$
V_{\rm EPR}\equiv(\Delta X)_{\rm inf}^2 (\Delta Y)_{\rm inf}^2<1.
$$
The variances are individually minimized with the choices
\begin{subequations}
\begin{eqnarray}
g_{X,\rm opt}&=& \frac{C_X}{2 S_{X2}}\\
g_{Y,\rm opt}&=&\frac{C_Y}{2 S_{Y2}}
\end{eqnarray}
\label{eq.qno.eprg1}
\end{subequations}
where $C_Y=\langle Y_{21}(\Omega) {Y_{22}}^*(\Omega)+{Y_{21}}^*(\Omega) Y_{22}(\Omega) \rangle/\delta(0).$ 
The minimum of the inference product is thus 
\begin{equation}
V_{\rm EPR}=
\frac{\left[S_{X1}S_{X2}-\frac{1}{4}C_X^2 \right]
\left[S_{Y1}S_{Y2}-\frac{1}{4}C_Y^2 \right]}{S_{X2}S_{Y2}}.
\label{eq.Vepr}
\end{equation}
We plot the inferred variance product in Fig. \ref{fig.epr1} as a function of the fundamental pump power. The variance product is less than 1, implying the outputs are EPR correlated, for pump power above about 30 mW. 

The presence of EPR correlations are a sufficient but not a necessary condition for entanglement of the two output beams. A necessary and sufficient condition for entanglement of Gaussian states is the inseparability of the density matrix describing the two output modes. This can be verified using the criterion of Duan et al.\cite{Duan00}.
In our notation this criterion can be written as 
\begin{eqnarray}
&&\min \left\{\left[\Delta(|a|X_{21}+\frac{1}{a} X_{22})\right]^2
 + \left[\Delta(|a|Y_{21}-\frac{1}{a} Y_{22})\right]^2
\right\}\nonumber
\\&&\hspace{3cm}<2\delta(0)\left(a^2+\frac{1}{a^2}\right),\nonumber
\end{eqnarray}
where $a$ is a real parameter. We have included the factor of 
$2\delta(0)$ on the right hand side  to account for our normalization of the commutators $[\hat X_{ij}(t),\hat Y_{kl}(t)]=2i\delta_{ik}\delta_{jl}$, which is different than that used in Ref. \cite{Duan00}. For the dual ported cavity the minimum is obtained for $a\simeq1$ so that the inseparability criterion can be written as 
\begin{eqnarray}
V_{\rm DGCZ}&=&\frac{1}{4}
\left(S_{X1}+S_{X2}+C_{X}+S_{Y1}+S_{Y2}-C_{Y} \right)<1.\nonumber\\
\label{eq.Vdgcz}
\end{eqnarray}
In order to facilitate comparison with $V_{\rm EPR}$ defined in Eq. (\ref{eq.Vepr}) we have included a factor of $1/4$ in the definition of 
$V_{\rm DGCZ}.$ Thus entanglement is indicated for both criteria by
$V<1.$ Figure \ref{fig.epr1} shows that $V_{\rm DGCZ}<1$ for all finite values of the pump power, and that  $V_{\rm DGCZ}<V_{\rm EPR}.$ These results verify that the harmonic fields at the two ouput ports are always entangled, but strong EPR correlations are only present when the pump power exceds a threshold value.

\section{EPR correlations on a beam splitter}
\label{sec:compare}

\begin{figure}[!t]
\begin{center}
\includegraphics[width=6.cm]{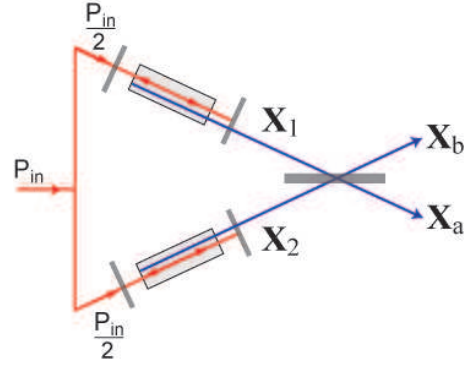}
\caption{(color online) Method for generating EPR correlations by mixing individually squeezed beams on a 50/50 beamsplitter.  }
\label{fig.entanglebs}
\end{center}
\end{figure}

An alternative approach to creating EPR correlations is to combine two individually squeezed beams on a beam splitter\cite{Furusawa98}, as shown in Fig. \ref{fig.entanglebs}. The two input beams are labeled with subscripts $1,2$ and the two output beams are labeled $a,b.$ For notational convenience we drop the first subscript labeling the harmonic frequency. We can choose the phase of the incident beams such that the harmonic field fluctuations transform as
\begin{subequations}
\begin{align}
 a_{a} &= \frac{ {a}_{1}-i {a}_{2}}{\sqrt{2}}\\
a_{b} &= \frac{ {a}_{1}+i{a}_{2}}{\sqrt{2}}
\end{align}
\end{subequations}
The quadrature fluctuations of the output fields are
\begin{subequations}
\begin{align}
 X_{a} = \frac{1}{\sqrt{2}}( X_{1}+ Y_{2})\\
 X_{b} = \frac{1}{\sqrt{2}}( X_{1}- Y_{2})\\
 Y_a = \frac{1}{\sqrt{2}}( Y_{1}- X_{2})\\
 Y_b = \frac{1}{\sqrt{2}}( Y_{1}+ X_{2}).
\end{align}
\label{eq.qno.xyab}
\end{subequations}

Following the same procedure as in the analysis of the dual ported cavity we define  the normalized inference variances at frequency $\Omega$ as
\begin{subequations}
\begin{eqnarray}
(\Delta X)_{\rm inf}^2&=& 
\frac{\langle |X_{a}(\Omega)-g_X X_{b}(\Omega)|^2\rangle}{\delta(0)}\\
(\Delta Y)_{\rm inf}^2&=& 
\frac{\langle |Y_{a}(\Omega)-g_Y Y_{b}(\Omega)|^2\rangle}{\delta(0)}.
\end{eqnarray}
\label{eq.qno.epr2c}
\end{subequations}
The optimum $g$ factors which minimized the inferred variances are
given by Eqs. (\ref{eq.qno.eprg1}) to be 
\begin{eqnarray}
g_{X,\rm opt}&=& \frac{\langle X_{a}(\Omega) {X_{b}}^*(\Omega)+{X_{a}}^*(\Omega) X_{b}(\Omega) \rangle}{2 \delta(0)S_{Xb}}=\frac{C_{Xab} }{2 S_{Xb}}\nonumber\\
g_{Y,\rm opt}&=&\frac{\langle Y_{a}(\Omega) {Y_{b}}^*(\Omega)+{Y_{a}}^*(\Omega) Y_{b}(\Omega) \rangle}{2 \delta(0) S_{Yb}}=\frac{C_{Yab} }{2 S_{Yb}}.\nonumber
\label{eq.qno.eprg2}
\end{eqnarray}

Hence, the inference product is 
$$
V_{\rm EPR}=
\frac{\left[S_{Xa}S_{Xb}-\frac{1}{4}C_{Xab}^2 \right]
\left[S_{Ya}S_{Yb}-\frac{1}{4}C_{Yab}^2 \right]}{S_{Xb}S_{Yb}}.
$$
Using Eqs. (\ref{eq.qno.xyab}) we find 
\begin{eqnarray}
S_{Xa}=S_{Xb}&=&\frac{1}{2}(S_{X1}+S_{Y2})\nonumber\\
S_{Ya}=S_{Yb}&=&\frac{1}{2}(S_{Y1}+S_{X2})\nonumber\\
C_{Xab}&=&
S_{X1}- S_{Y2}\nonumber\\
C_{Yab}&=&
S_{Y1}-S_{X2}.\nonumber
\end{eqnarray}
When the input beams $1,2$ are generated in equivalent resonators we have $S_{X1}=S_{X2}$, 
$S_{Y1}=S_{Y2}$ and the  EPR correlation  reduces 
to 
$$
V_{\rm EPR}=\frac{4 S_{X1}^2S_{Y1}^2}{(S_{X1}+S_{Y1})^2},
$$
where the condition for an EPR paradox is $V_{\rm EPR}<1.$
As seen in Fig. \ref{fig:squeeze} second harmonic generation results in nonideal squeezed states with $S_{X1}S_{Y1}=p>1.$ We can then write $V_{\rm EPR}=4 S_{X1}^2p^2/(S_{X1}^2 +p)^2.$

\begin{figure}[!t]
\begin{center}
\includegraphics[width=8.cm]{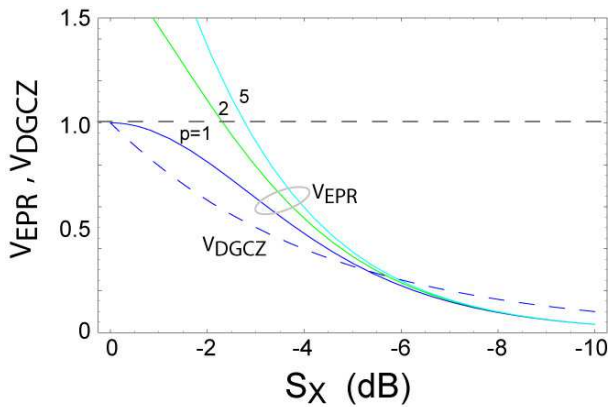}
\caption{(color online) EPR (solid lines)  and inseparability (dashed line) correlation coefficients for nonideal squeezed beams combined on a beamsplitter.  }
\label{fig.epridealfactor}
\end{center}
\end{figure}

The EPR correlation is shown in Fig. \ref{fig.epridealfactor} for several values of $p.$  We see that for strong amplitude squeezing the EPR correlation is quite insensitive to the degree of excess noise in the phase quadrature. On the other hand for moderate levels of squeezing, less than about 4 dB, 
$V_{\rm EPR}$ increases significantly with $p.$ Referring to Fig. \ref{fig:squeeze} we see that $P_{\rm in}=0.5~\rm W$ gives $S_X= -2 ~\rm dB$ and $p=2.2$. For these values the EPR correlation obtained by mixing two individually squeezed beams on a beamsplitter is $V_{\rm EPR}=1.1$, so there is no EPR paradox. On the other hand for $P_{\rm in}=0.5 ~\rm W$ the dual ported resonator gives 
$V_{\rm EPR}=0.7.$ Thus the dual ported resonator is able to generate 
much stronger EPR correlations even though the output beams have a lower level of squeezing than is required using a beamsplitter to mix two squeezed sources. 

When combining  two equivalent squeezed sources on a beamsplitter the DGCZ criterion as given by Eq. (\ref{eq.Vdgcz}) takes on the simple form $V_{\rm DGCZ}=S_X,$ which is independent of the parameter $p.$ Thus, as seen in Fig. \ref{fig.epridealfactor}, nonseparable beams
can always be created by beamsplitter mixing, even using nonideal squeezed sources.

\begin{figure}[!t]
\begin{center}
\includegraphics[width=8.cm]{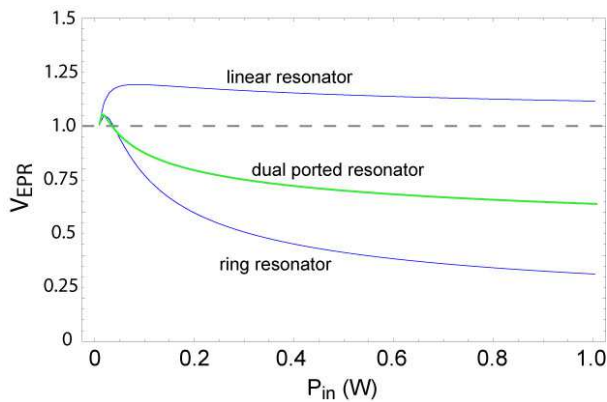}
\caption{(color online) Correlation coefficient $V_{EPR}$ produced by mixing outputs from  dual and single port resonators.  The correlation is calculated as a function of the total input power, so that in the case of two linear or ring resonators, each one is pumped by $P_{\rm in}/2$ as shown in Fig. \ref{fig.entanglebs}.  }
\label{fig.eprbs}
\end{center}
\end{figure}

We can also compare the EPR correlation generated in the dual ported resonator directly with beamsplitter mixing as a function of total pump power. As seen in Fig. \ref{fig.eprbs} the value of $V_{\rm EPR}$ generated using two sources and a beamsplitter depends strongly on the characteristics of the source. The upper curve labeled linear resonator shows the results obtained by using one of the outputs 
from a dual ported resonators with the parameters used in Figs. \ref{fig:squeeze} to \ref{fig.epr1}. This is clearly suboptimal as only half of the generated harmonic light is being used, and no EPR paradox is seen for moderate pump powers in this case. The lower curve labeled ring resonator corresponds to an optimized ring resonator which has the same parameters as for the dual ported resonator except $E_{NL2}=0.$
This resonator generates much stronger squeezing, and also a smaller 
$V_{\rm EPR}$ than the dual ported resonator. We defer until the next section a discussion of  the optimal approach to generating EPR correlations using SHG.

\section{Discussion}
\label{sec:discussion}

We have shown that SHG in a resonator with two output ports can be used to generate output beams that exhibit nonclassical intensity correlations, as well as entanglement. The analysis uses a propagation model for the quantum  fluctuations that goes beyond the usual mean field approximation. We have shown results for the zero frequency noise fluctuations. The nonclassical correlations and entanglement will degrade as the frequency considered is increased, with the 
frequency dependence following the usual Lorentizan form for a nonlinear cavity\cite{Paschotta94}. The use of SHG to create entangled beams, as opposed to the more commonly employed method of parametric oscillation or amplification, may be advantageous in that it provides additional flexibility in the choice of spectral region for the entangled beams. It is  also possible to combine two individually squeezed beams with a 50-50 beamsplitter to create entanglement\cite{Furusawa98}. The question therefore arises as to which method, the dual ported resonator or two separate resonators, is most efficient at creating usable entanglement. This question has been considered for the case of degenerate vs. nondegenerate parametric oscillation in Ref. \cite{Dechoum04} where it was shown that a single nondegenerate oscillator was generally preferred. 

Referring to Fig. \ref{fig.eprbs} we see that the EPR correlations strength for a given available pump power obtainable with the dual ported resonator lies in between the results obtained by combining the outputs of two separate dual ported, or single port ring resonators. 
The results presented were obtained for  resonator parameters given in the text. Although differences occur for different parameter choices we believe the results presented are representative. Given available nonlinear crystals which fix the value of $E_{NL}$ the most important parameters that can be varied are the internal losses and input coupler transmissions. Results have been given for round trip passive losses of $L_{13}+L_{14}=1\%$. Reducing this further gives larger nonclassical effects but seems unrealistic for current experiments. We have also chosen a low input coupler transmission of $T_{11}=1\%.$ While this does not give the maximum possible harmonic conversion we have found that it is close to optimum for generating squeezing. 

Although Fig. \ref{fig.eprbs} appears to show  that, as far as  EPR correlations are concerned, it is more efficient to combine two separate resonators, this conclusion may not be warranted. The single output resonator which generates the strongest EPR correlation in Fig. \ref{fig.eprbs} does so by producing large amounts of squeezing. For the parameters used in the text and $P_{\rm in}/2=0.5~\rm W$ the ring resonator is predicted to give  about 6 dB of amplitude squeezing into a single harmonic output. 
On the other hand the largest amount of amplitude squeezing ever reported in a SHG experiment was, to the best of our knowledge, measured to be 2.4 dB, with an inferred squeezing 
of 5.2 dB\cite{Tsuchida95}. It may therefore not be possible to achieve the level of squeezing predicted in the theoretical model. The difficulties include parasitic effects such as blue light induced infrared absorption that become prominent when large amounts of the harmonic field are generated\cite{Mabuchi94}. There are also more fundamental limitations present in high conversion efficiency SHG
due to the excitation of competing parametric processes which have been shown to limit the level of harmonic squeezing that is attainable\cite{White97}. 

Given these considerations the use of a dual ported resonator which 
generates a strong EPR correlation even though the squeezing level and the power of each beam are not large may be advantageous compared to combining separate SHG resonators on a beam splitter. Finally, we note that the dual ported configuration of Fig. \ref{fig:linear} is also attractive in terms of experimental simplicity, compared to a two resonator plus beamsplitter arrangement.

\begin{acknowledgments}
This work was supported by NSF grant ECS-0533472.
\end{acknowledgments}

\appendix

\section{\label{sec:appA} Effective interaction strength}

In order to use Eqs. (\ref{eq.outputs})  to calculate output spectra we must first evaluate the interaction strength in each crystal. Because the fundamental is partially converted to the second harmonic in the first pass through the crystal  the interaction in the second pass  will be slightly weaker. 

\begin{figure*}[!t]
\begin{center}
\includegraphics[width=14.cm]{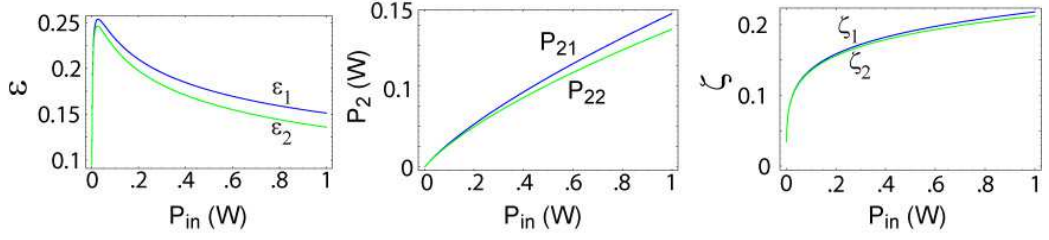}
\caption{(color online) Power conversion efficiency, second harmonic power and normalized propagation length, in the dual output cavity.}
\label{fig.epsandzeta}
\end{center}
\end{figure*}

The cavity geometry is shown in Fig. \ref{fig:unfolded}.  After some simple algebra we find for the conversion efficiency of crystal 1
\begin{widetext}\begin{equation}
\sqrt{\epsilon_1}=\frac{4 T_{11}\sqrt{E_{NL1}P_{\rm in}}}
{\left[2-\sqrt{1-T_{11}}\sqrt{1-T_{12}}\left(2-L_{13}-L_{14}-\sqrt{\epsilon_1E_{NL1}P_{\rm in}} -\sqrt{\epsilon_2E_{NL2}P_{\rm in}} \right)\right]^2}.
\label{eq.eps1}
\end{equation}
\end{widetext}
We have introduced the conversion efficiencies $\epsilon_1=P_{21}/P_{\rm in},$ $\epsilon_2=P_{22}/P_{\rm in},$ $P_{21}, P_{22}$ are the harmonic output powers after the first and second crystal passes, $P_{\rm in}$ is the fundamental pump power external to the cavity,  and the other parameters are defined in Sec. \ref{sec:SHG}.
  When $E_{NL2}=L_{14}=T_{12}=0$ Eq. (\ref{eq.eps1}) reduces to the known result\cite{Polzik91} for a single crystal ring cavity.

In the two crystal case the conversion efficiencies are related by 
\begin{equation}
\epsilon_2=\frac{E_{NL2}}{E_{NL1}}(1-T_{12})^2(1-L_{13})^2\epsilon_1(1-\sqrt{\epsilon_1 E_{NL1}P_{in}})^2.
\label{eq.eps2}
\end{equation}
Using this result in (\ref{eq.eps1}) we get a closed expression for $\epsilon_1$ that can be solved numerically. The normalized propagation lengths are then given by
\begin{subequations}
\begin{eqnarray}
\zeta_1&=&\sqrt{\frac{n_1}{n_2}\sqrt{\epsilon_1E_{NL1}P_{in}}}\\
\zeta_2&=&\sqrt{\frac{n_1}{n_2}\sqrt{\epsilon_2E_{NL2}P_{in}}}.
\end{eqnarray}
\label{eq.zeta12}
\end{subequations}

Numerical examples using the same parameters as in Sec. \ref{sec:SHG} ($E_{NL1}=E_{NL2}=.015~\rm W^{-1},$ $n_1=n_2=2.2,$ 
$T_{11}=.01,$ $T_{12}=0,$  $L_{13}=L_{14}=.005$) are shown in Fig.  \ref{fig.epsandzeta}. The conversion efficiencies, second harmonic power, and propagation lengths are shown in Fig. \ref{fig.epsandzeta}
as a function of the input pump power. The parameters were chosen to give strong nonclassical correlations, but are not optimal for power conversion since $\epsilon_1, \epsilon_2$ peak at quite low power, and the cavity is undercoupled at higher input powers. 
We see that the normalized propagation lengths $\zeta_1, \zeta_2 \ll 1$ so the linearized analysis used in the paper is reliable. In addition the fractional pump depletion due to a single pass of the intracavity field with power $P_c$ through the crystal is given by $\Delta P_c/P_c=\sqrt{\epsilon_1 E_{NL1}P_{\rm in}}.$ For the parameters used the fractional depletion is always less than 5\%, so the Boyd-Kleinman analysis which is based on the assumption of an unaltered spatial form for the fundamental field is a good approximation.

\section{\label{sec:appB}Coefficients of output quadratures }

The coefficients appearing in Eqs. (\ref{eq.outputs2}) are 

\begin{widetext}

\begin{eqnarray}
f_{11}&=&\frac{e^{\imath \Omega/\nu_{c1}}\sqrt{T_{11}}N_{21}(\zeta_1)}{F}\nonumber\\
f_{13}&=&-\frac{e^{\imath \Omega/\nu_{c1}}\sqrt{1-T_{11}}\sqrt{L_{13}}\sqrt{1-L_{14}}N_{21}(\zeta_1)N_{11}(\zeta_2)}{F}\nonumber\\
f_{14}&=&-\frac{e^{\imath \Omega/\nu_{c1}}\sqrt{1-T_{11}}    \sqrt{L_{14}} N_{21} (\zeta_1)} {F}\nonumber\\
f_{21}&=&\frac{N_{22}(\zeta_1)-e^{\imath \Omega/\nu_{c1}}
\sqrt{1-T_{11}}\sqrt{1-L_{13}}\sqrt{1-L_{14}}N_{11}(\zeta_2)[N_{11}(\zeta_1)N_{22}(\zeta_1)-N_{12}(\zeta_1)N_{21}(\zeta_1)]}{F}\nonumber\\
f_{22}&=&\frac{e^{\imath \Omega/\nu_{c1}}\sqrt{1-T_{11}}\sqrt{1-L_{14}}N_{21}(\zeta_1)N_{12}(\zeta_2)}{F}
\end{eqnarray}

\begin{eqnarray}
g_{11}&=&\frac{e^{\imath \Omega/\nu_{c1}}\sqrt{T_{11}}N_{43}(\zeta_1)}{G}\nonumber\\
g_{13}&=&-\frac{e^{\imath \Omega/\nu_{c1}}\sqrt{1-T_{11}}\sqrt{L_{13}}\sqrt{1-L_{14}}N_{43}(\zeta_1)N_{33}(\zeta_2)}{G}\nonumber\\
g_{14}&=&-\frac{e^{\imath \Omega/\nu_{c1}}\sqrt{1-T_{11}}    \sqrt{L_{14}} N_{43} (\zeta_1)} {G}\nonumber\\
g_{21}&=&\frac{N_{44}(\zeta_1)-e^{\imath \Omega/\nu_{c1}}
\sqrt{1-T_{11}}\sqrt{1-L_{13}}\sqrt{1-L_{14}}N_{33}(\zeta_2)[N_{33}(\zeta_1)N_{44}(\zeta_1)-N_{34}(\zeta_1)N_{43}(\zeta_1)]}{G}\nonumber\\
g_{22}&=&\frac{e^{\imath \Omega/\nu_{c1}}\sqrt{1-T_{11}}\sqrt{1-L_{14}}N_{43}(\zeta_1)N_{34}(\zeta_2)}{G}
\end{eqnarray}

\begin{eqnarray}
h_{11}&=&\frac{e^{\imath \Omega/\nu_{c1}}\sqrt{T_{11}}\sqrt{1-L_{13}}N_{11}(\zeta_1)N_{21}(\zeta_2)}{F}\nonumber\\
h_{13}&=&-\frac{\sqrt{L_{13}}N_{21}(\zeta_2)}{F}\nonumber\\
h_{14}&=&-\frac{e^{\imath \Omega/\nu_{c1}}\sqrt{1-T_{11}}    \sqrt{1-L_{13}}\sqrt{L_{14}} 
N_{11}(\zeta_1)N_{21} (\zeta_2)} {F}\nonumber\\
h_{21}&=&\frac{\sqrt{1-L_{13}}N_{12}(\zeta_1))N_{21}(\zeta_2)}{F}\nonumber\\
h_{22}&=&\frac{N_{22}(\zeta_2)-e^{\imath \Omega/\nu_{c1}}\sqrt{1-T_{11}}\sqrt{1-L_{13}}\sqrt{1-L_{14}}N_{11}(\zeta_1)[N_{11}(\zeta_2)N_{22}(\zeta_2)-N_{12}(\zeta_2)N_{21}(\zeta_2)]}{F}
\end{eqnarray}

\begin{eqnarray}
j_{11}&=&\frac{e^{\imath \Omega/\nu_{c1}}\sqrt{T_{11}}\sqrt{1-L_{13}}N_{33}(\zeta_1)N_{43}(\zeta_2)}{G}\nonumber\\
j_{13}&=&-\frac{\sqrt{L_{11}}N_{43}(\zeta_2)}{G}\nonumber\\
j_{14}&=&-\frac{e^{\imath \Omega/\nu_{c1}}\sqrt{1-T_{11}}    \sqrt{1-L_{13}}\sqrt{L_{14}} 
N_{33}(\zeta_1)N_{43} (\zeta_2)} {G}\nonumber\\
j_{21}&=&\frac{\sqrt{1-L_{13}}N_{34}(\zeta_1)N_{43}(\zeta_2)}{G}\nonumber\\
j_{22}&=&\frac{N_{44}(\zeta_2)-e^{\imath \Omega/\nu_{c1}}\sqrt{1-T_{11}}\sqrt{1-L_{13}}\sqrt{1-L_{14}}N_{33}(\zeta_1)[N_{33}(\zeta_2)N_{44}(\zeta_2)-N_{34}(\zeta_2)N_{43}(\zeta_2)]}{G}\nonumber\\
\end{eqnarray}
\end{widetext}

In the above expressions, we have introduced the definitions
\begin{align*}
F &= 1-e^{\imath \Omega/\nu_{c1}}\sqrt{1-T_{11}}\sqrt{1-L_{13}}\sqrt{1-L_{14}}N_{11}(\zeta_1)N_{11}(\zeta_2),\\
G &= 1-e^{\imath \Omega/\nu_{c1}}\sqrt{1-T_{11}}\sqrt{1-L_{13}}\sqrt{1-L_{14}}N_{33}(\zeta_1)N_{33}(\zeta_2).
\end{align*}
The propagation matrix elements are\cite{Ou94,Li94} 
\begin{eqnarray}
N_{11}(\zeta)&=&\frac{1-\zeta\tanh\zeta}{\cosh\zeta},~~
N_{12}(\zeta)=-\sqrt2\frac{\tanh\zeta}{ \cosh\zeta},\nonumber\\
N_{21}(\zeta)&=&\frac{1}{\sqrt2}(\tanh\zeta+\zeta{\rm sech}^2\zeta),~~
N_{22}(\zeta)={\rm sech}^2\zeta,\nonumber\\
N_{33}(\zeta)&=&{\rm sech}\zeta,~~
N_{34}(\zeta)=-\frac{1}{\sqrt2}(\sinh\zeta+\zeta{\rm sech}\zeta),\nonumber\\
N_{43}(\zeta)&=&\sqrt2\tanh\zeta,~~
N_{44}(\zeta)=1-\zeta\tanh\zeta.\nonumber
\end{eqnarray}

%\bibliography{pub}

\end{document}